\newcommand{\be}{\begin{equation}}
\newcommand{\bea}{\begin{IEEEeqnarray}}
\newcommand{\ee}{\end{equation}}
\newcommand{\eea}{\end{IEEEeqnarray}}
\newcommand{\nn}{\nonumber}

\newcommand{\qa}{\alpha}
\newcommand{\qb}{\beta}

\newcommand{\qD}{\Delta}
\newcommand{\qe}{\varepsilon}

\newcommand{\ql}{\lambda}

\newcommand{\qr}{\rho}
\newcommand{\qs}{\sigma}

\newcommand{\qt}{\tau}
\newcommand{\qf}{\varphi}

\newcommand{\qo}{\omega}
\newcommand{\qO}{\Omega}

\newcommand{\Tr}{{\rm Tr}\,}

\newcommand{\dagg}{^{\dag}}

\newcommand{\fr}[2]{{\textstyle \frac{#1}{#2}}}

\newcommand{\EE}{{\mathbb E}}

\newcommand{\one}{\mathbb{1}}
\newcommand{\sH}{{\sf H}}
\newcommand{\bits}{ \{0,1\} }
\newcommand{\Hmin}{{\sf H}_{\rm min}}
\newcommand{\cA}{{\mathcal A}}

\newcommand{\cC}{{\mathcal C}}
\newcommand{\cD}{{\mathcal D}}

\newcommand{\cH}{{\mathcal H}}

\newcommand{\cK}{{\mathcal K}}

\newcommand{\cO}{{\mathcal O}}

\newcommand{\cS}{{\mathcal S}}

\newcommand{\cX}{{\mathcal X}}

\newcommand{\pr}{{\rm Pr}}

\newcommand{\isdef}{\stackrel{\rm def}{=}}

\newcommand{\ket}[1]{| #1 \rangle}
\newcommand{\bra}[1]{\langle #1 |}

\documentclass[lettersize,journal]{IEEEtran}
\usepackage{amsmath,amsfonts,cuted}
\usepackage{algorithmic}
\usepackage{algorithm}
\usepackage{array}
\usepackage[caption=false,font=normalsize,labelfont=sf,textfont=sf]{subfig}
\usepackage{textcomp}
\usepackage{stfloats}
\usepackage{url}
\usepackage{verbatim}
\usepackage{graphicx}
\usepackage{cite}
\usepackage{bbold}
\usepackage{makecell}
\usepackage{enumitem}
\newtheorem{theorem}{Theorem}[section]

\newtheorem{lemma}[theorem]{Lemma}
\newtheorem{definition}[theorem]{Definition}

\begin{document}
\title{Entropically secure encryption\\ with faster key expansion}

\author{Mehmet H\"{u}seyin Temel and Boris \v{S}kori\'{c}
\thanks{Part of this work was supported by the Dutch Startimpuls NAQT KAT-2 and NGF Quantum Delta NL KAT-2.}
}

\markboth{}%
{}

\IEEEpubid{}

\maketitle

\begin{abstract}
Entropically secure encryption is a way to encrypt a large plaintext with a small key
and still have information-theoretic security, thus in a certain sense circumventing
Shannon's result that perfect encryption requires the key to be at least as long as
the entropy of the plaintext.
Entropically secure encryption is not perfect, and it works only if a lower bound is 
known on the entropy of the plaintext.
The typical implementation is to expand the short key to the size of the plaintext,
e.g. by multiplication with a public random string,
and then use one-time pad encryption.
This works in the classical as well as the quantum setting. In this paper, we introduce a new key expansion method that is faster than
existing ones. We prove that it achieves the same security.
The speed gain is most notable when the key length is a sizeable fraction of 
the message length. 
In particular, a factor of 2 is gained in the case of approximate randomization of quantum states.
\end{abstract}


\section{Introduction}
\subsection{Entropic Security}
\IEEEPARstart{A}{n} encryption scheme is called {\em perfect} if the ciphertext reveals no information whatsoever about the plaintext. 
For perfect encryption of {\em classical} plaintexts, the length of the key needs to be at least the entropy of the plaintext, and
the simplest cipher is the One-Time Pad (OTP) or Vernam cipher \cite{Vernam1918}.
In the {\em quantum} setting, perfect encryption of an $n$-qubit plaintext state requires a key length of
$2n$ bits and the simplest cipher achieving this kind of encryption is the Quantum One-Time Pad (QOTP) \cite{AMTW2000, BR2003, Leung2002}.

If one does not aim for {\em perfect} security, it is possible to get information-theoretic guarantees about the encryption
even with shorter keys, as long as a lower bound is known on the min-entropy of the plaintext. 
The notion of $(t,\qe)$-entropic security has been introduced \cite{russell2002fool,dodis2005entropic}, stating that the adversary's advantage in guessing any function of the plaintext is upper bounded by $\qe$ if the min-entropy of the plaintext 
(conditioned on Eve's side information) is at least~$t$.
It can be seen as an information-theoretic version of semantic security. It has been shown that $(t,\qe)$-entropically secure encryption of
an $n$-(qu)bit plaintext can be achieved with key length $n-t+2\log\fr1\qe$
\cite{dodis2005entropic,Des2009,DD2010}.
In the quantum case, the $t$ can become negative when Eve's quantum memory is entangled with the plaintext state.

The standard way to perform entropically secure encryption is to expand the short key to a pseudorandom string which is then used as the key for (Q)OTP encryption.
The key expansion is done either using small-bias spaces or by universal hashing with a public random string.

\subsection{Contribution and Outline}
We introduce a new key expansion method for entropically secure encryption, both classical and quantum.
The main idea is to {\em append} a pseudorandom string $f(k)$ to the short key $k$, 
instead of creating an entirely new string from~$k$.
For the computation of $f(k)$, we use finite-field multiplication with a public random string.
\begin{itemize}[leftmargin=3mm,itemsep=0mm]
\item
Our key expansion is faster than all previous constructions while achieving the shortest known
key length $n-t+2\log\fr1\qe$.
In particular, a factor of 2 in speed is gained in the unentangled
quantum case without further assumptions on Eve and the entropy of the plaintext.
\item Our security proof in the quantum case is a bit more straightforward than  \cite{DD2010}, as we avoid expanding states in the Pauli basis.
\end{itemize}
The outline of the paper is as follows.
We discuss related work on entropic security in Section~\ref{sec:related}.
We present the relevant definitions and lemmas from the literature in Section~\ref{sec:prelim}.
In Section~\ref{sec:classical}, we present our classical scheme and its security proof.
Then in Section~\ref{sec:quantum}, we do the same for the quantum scheme, additionally giving an analysis of 
the computational complexity of the key expansion, especially in the unentangled case in Section~\ref{sec:expansion}.
Finally, in Section~\ref{sec:discussion}, we comment on possible improvements.

\section{Related work}
\label{sec:related}

\subsection{Entropic Security for Classical Plaintext}

The entropic security notion and its definition for encryption were coined first by Russell and Wang \cite{russell2002fool}. 
They showed that it is possible to encrypt a high-entropy $n$-bit plaintext using a key shorter than $n$ such that
an attacker has less than $\qe$ advantage in predicting {\em predicates} of the plaintext.
Such schemes are called Entropically Secure Encryption (ESE) schemes. 
They provided an ESE with key length $n-t+3\log\fr1\varepsilon +\cO(1)$, where $t$ is the min-entropy of the plaintext. 

Dodis and Smith \cite{dodis2005entropic} gave a stronger definition by considering {\em all functions} of the plaintext instead of only predicates. 
They also showed the equivalence between ESE and indistinguishability (in terms of statistical distance) of ciphertexts.
They introduced two simple constructions,
one of which uses XOR-universal hash functions and improves the key length to 
$n-t+2\log\fr1\varepsilon + \cO(1)$.
They proved that the key length of any ESE needs to be at least $n-t$ bits. 

Fehr and Schaffner \cite{fehr2008randomness} introduced a classical indistinguishable encryption scheme secure against quantum adversaries. 
It has key length $n-t+2\log n +2\log\fr1\varepsilon+\cO(1)$,
where $t$ is the collision entropy of the (classical) plaintext given Eve's quantum side information.
Different from \cite{russell2002fool, dodis2005entropic} they used collision entropy in the security definition
instead of min-entropy.

\subsection{Entropic Security in the Quantum Setting}

In order to perfectly encrypt any $n$-qubit state, the necessary and sufficient key length is $2n$ bits \cite{AMTW2000,BR2003,Leung2002}. 
In its simplest form, quantum one-time pad (QOTP) encryption and decryption work by applying to each individual qubit a Pauli operation from the set $\{\one,\qs_x,\qs_y,\qs_z\}$. 
The choice of Pauli operations constitutes the key. 
For someone who does not know this key, the state after encryption equals the fully mixed state regardless of the plaintext state. 

Entropic security has been generalized to the fully quantum setting where both the plaintext and ciphertext are quantum states. 
Desrosiers \cite{Des2009} introduced definitions of entropic security and entropic indistinguishability for quantum ciphers.
Similar to the classical setting, these definitions are equivalent up to parameter changes.
He also introduced a scheme with a key length of $n-t+2\log\fr1\varepsilon$ using a similar key expansion method as \cite{dodis2005entropic}.
Here $t$ is the min-entropy of the plaintext quantum state.
The analysis in \cite{Des2009} applies only if Eve is not entangled with the plaintext.
Desrosiers and Dupuis \cite{DD2010} generalized the analysis, with conditional quantum min-entropy
as defined by Renner \cite{renner2008security},
and showed that the results hold even with entanglement.\footnote{
The conditional quantum min-entropy of an $n$-qubit system ranges between $-n$ and $n$. 
A maximally entangled state has conditional min-entropy $t=-n$.
} 
They also proved a minimum required key length of $n-t-1$.

\subsection{Approximate Randomization}

The quantum setting without entanglement ($t\geq 0$) is a special case that has received
a lot of attention by itself, as it occurs naturally:
It corresponds to the situation where Alice prepares a plaintext state and encrypts it. 
As long as the plaintext state is 
generated entirely by Alice (as opposed to being received by Alice as part of a larger protocol),
Eve is not entangled with it.
In the literature, this special case goes by the name {\em approximate randomization} or approximate quantum encryption.
Hayden et al. \cite{HLSW2004} showed that approximate randomization is possible with a key length of $n+\log n + 2\log\fr1\qe$ by using sets of random unitaries.\footnote{
They also provided a result for the $\infty$-norm, with key length $n+\log n + 2\log\fr1\qe + \log134$; unitaries are drawn from the
Haar measure. This was later improved to $n+ 2\log\fr1\qe + \log150$ by Aubrun \cite{Aubrun2009}.} 
Ambainis and Smith \cite{ambainis2004small} introduced far more efficient schemes that work with a pseudorandom sequence which selects Pauli operators as in the QOTP. In one of them, they expanded the key using small-bias sets and achieve key length $ n+2\log n+2\log\fr1\qe$. This scheme is length-preserving, i.e.\;the cipherstate consists of $n$ qubits. In another construction, they expanded the key by multiplying it with a random binary string of length $2n$; this string becomes part of the cipherstate.
The key length is reduced to $n+2\log\fr1\qe$.
Dickinson and Nayak \cite{DN2006} improved the small-bias based scheme of \cite{ambainis2004small} and achieved key length $n+2\log\fr1\qe+4$. \v{S}kori\'{c} and de Vries \cite{SdV2017} described a pseudorandom QOTP scheme that has key length $n+2\log\fr1\qe$, but they need an exponentially large common random string to be stored.
Table~\ref{table:1} shows an overview of these results.

\begin{table*}[h!]
\caption{Results on approximate randomization of $n$ qubits \label{table:1}}
\begin{center}
\begin{tabular}{ |c||c|>{\centering\arraybackslash}m{2.5cm}|m{10cm}|}
\hline
  & {\it Key length} $\ell$ & {\it Ciphertext length} & {\it Randomization process} \\
 \hline
 \hline
 \cite{HLSW2004} & $n+\log n + 2\log\fr1\qe$& $n$ qubits & 
Random unitaries (non-Haar, e.g. Pauli)
 \\
 \hline
 \cite{ambainis2004small} & $ n+2\log n+2\log\fr1\qe$ & $n$ qubits & 
Pseudorandom QOTP based on small-bias sets. 
Key expansion takes  $\cO(n^2)$ operations.
\\
 \hline
 \cite{ambainis2004small} & $n+2\log\fr1\qe$ & $n$ qubits + $2n$ bits & 
Pseudorandom QOTP based on multiplication in GF$(2^{2n})$.
Key expansion takes $\approx 6n\log_3 n$ operations.
\\
 \hline
 \cite{DN2006} & $n+2\log\fr1\qe +4$& $n$ qubits & 
Pseudorandom QOTP based on small-bias spaces. 
Key expansion takes ${\cal O}(n^2 \log n)$ operations.
\\
 \hline
 \cite{SdV2017} & $n+2\log\fr1\qe$ & $n$ qubits & 
Pseudorandom QOTP based on huge Common Reference String.
\\
 \hline
 This work & $n+2\log\fr1\qe$ & $n$ qubits + $2n$ bits & 
Pseudorandom QOTP based on affine function in  GF($2^\ell$).
Key expansion takes $\approx 3n\log_3 n$ operations.
\\
 \hline
\end{tabular}
\\
\it The security is in terms of the trace distance: $\| {\rm cipherstate}-\mbox{\rm fully mixed state}  \|_1\leq\qe$.
The listed complexity for the finite field \\
multiplications is based on
the fastest known implementation and shows only the number of \textit{AND} operations.
(See Section~\ref{sec:expansion})
\end{center}
\end{table*}

The `$\qe$-close to fully mixed' property can be expressed as a distance with respect to different norms, e.g. the $1$-norm (trace norm) or the $\infty$-norm (maximum absolute eigenvalue). In this paper, we consider only the $1$-norm, since it expresses the distinguishability of states and it is a universally composable measure of security \cite{Can2001,ben2005universal,fehr2009composing}.

\section{Preliminaries}
\label{sec:prelim}

\subsection{Notation; Entropic Quantities}
\label{sec:notation}

Random variables are written in uppercase and their realizations in lowercase.
The statistical distance between random variables $X,Y\in\cX$ is given by
$\Delta(X,Y)=\fr12\sum_{a\in\cX}|\pr[X=a]-\pr[Y=a]|$.
Expectation over a random variable $X$ is written as~$\EE_x$.
The R\'{e}nyi entropy of order $\qa$ (with $\qa\neq 1$) is defined as 
$\sH_\qa(X)= (1-\qa)^{-1}\log \sum_{x}\pr[X=x]^\qa$.
Special cases are the collision entropy
$\sH_2(X)=-\log(\sum_{x}\pr[X=x]^2)$ and the min-entropy
$\Hmin(X)=\sH_\infty(X)=-\log(\max_x \pr[X=x])$.

We denote the space of density matrices on Hilbert space $\cH$ as $\cD(\cH)$.
The single-qubit Hilbert space is denoted as~$\cH_2$.
A bipartite state comprising subsystems `A' and `B' is written as 
$\qr^{\rm AB}\in\cD(\cH_{\rm A}\otimes \cH_{\rm B})$.
The state of a subsystem is obtained by taking the partial trace over the other subsystem,
e.g.\;$\qr^{\rm A}=\Tr_{\!\rm B}\,\qr^{\rm AB}$.
The identity operator on $\cH$ is denoted by $\one_\cH$; we will simply write $\one$ when 
the Hilbert space is clear from the context.
Similarly, we write $\qt^\cH$ for the fully mixed state $\one_\cH/{\rm dim}(\cH)$, often omitting the superscript.
Let $M$ be an operator with eigenvalues~$\ql_i$.
The Schatten $1$-norm of $M$ is given by 
$ \|  M \|_1
=\Tr\sqrt{M\dagg M}=\sum_i |\ql_i|$.
The trace distance between states $\qr,\qs$ 
is $ \|  \qr-\qs \|_{\rm tr}= \fr12\|  \qr-\qs \|_1$.

In the literature, there are multiple definitions of quantum conditional R\'{e}nyi entropy.
We will work with the definition of \cite{muller2013quantum}, which is based on 
sandwiched relative entropy, and which
has been shown in \cite{tomamichel2014relating} to possess the most favorable properties.
Let $\qr,\qs\geq 0$ with $\qr\neq 0$, and $\qa\in(0,1)\cup(1,\infty)$.
The sandwiched quantum R\'enyi relative entropy is given by
\bea{c}
	D_{\qa}(\qr||\qs) \isdef 
	\frac{1}{\qa-1}\log\Tr\Big[\Big( \qs^{\fr{1-\qa}{2\qa}} \qr\, \qs^{\fr{1-\qa}{2\qa}} \Big)^{\qa} \Big]
\label{RenyiDiv}
\eea
if $\qs\gg\qr\,\vee\, (\qa<1 \wedge \qr\not\perp\qs)$, and otherwise $D_{\qa}(\qr||\qs)=\infty$.
Here $\qs\gg\qr$ denotes that $\qs$ dominates $\qr$, i.e.\,that the kernel of $\qs$ is contained in the kernel of~$\qr$.
For $\qa\in(0,1)\cup(1,\infty)$ and $\qr^{\rm AE}\in\cD(\cH_A\otimes \cH_E)$, the conditional quantum
R\'{e}nyi entropy is given by
\be
    \sH_{\qa}(A|E)_\qr \isdef -\inf_{\qs \in \cD(\cH_E)} D_\qa(\qr^{\rm AE}|| \one_A\otimes\qs).
\label{condRenyi}
\ee
In the limit $\qa\to\infty$ this reproduces the conditional R\'{e}nyi entropy as introduced by Renner \cite{renner2008security},
\be
    \Hmin(A|E)_\qr =\sup_{\qs\in\cD(\cH_E)} -\log \min \{\ql \,|\, \ql \one_A\otimes \qs -\qr^{\rm AE}\geq 0\}.
\ee

\subsection{Security Definitions and Useful Lemmas}
\label{sec:secdef}

\begin{definition}[Classical encryption scheme]
\label{def:classicencryption}
A classical encryption scheme for a message space $\cX$ with key space $\cK$ and ciphertext space $\cC$ consists of three algorithms 
$({\sf Gen, Enc, Dec})$. 
{\sf Gen} is a probabilistic algorithm that outputs a key $k\in\cK$. 
The {\sf Enc}: $\cK\times\cX\to\cC$ is a possibly randomized algorithm that takes a key $k$ and a message $x\in\cX$ and outputs a ciphertext $c\in\cC$. 
The {\sf Dec}: $\cK\times\cC\to\cX$ is the decryption algorithm that takes a key $k\in\cK$ and a ciphertext $c\in\cC$ as inputs and outputs a message $x\in\cX$. 
It must hold that $\forall_{k\in\cK,x\in\cX}{\sf Dec}(k,{\sf Enc}(k, x)) = x$.
\end{definition}

Entropic security of an encryption scheme is formulated as
the property that the encryption {\sf Enc} hides all functions of the plaintext~$X$.
`Hiding' means that an adversary $\cA$ who sees the ciphertext has no advantage
in guessing function values $f(X)$ over an adversary $\cA'$ who does not observe the ciphertext.
Here the function $f$ is fixed {\em before} $\cA$ observes the ciphertext.
This is phrased in \cite{dodis2005entropic} as: `$Y$ leaks no {\em a priori} information about $X$'.\\
\begin{definition}[Hiding all functions]
\label{def:hiding}
A probabilistic map $Y$ is said to hide all functions of $X$ with leakage $\qe$ if
\be
	\forall_\cA\exists_{\cA'}\forall_f \quad \Big| 
	\pr[\cA(Y(X))=f(X)] - \pr[\cA'()=f(X)]
	\Big| \leq\qe.
\ee
\end{definition}

\begin{definition}[Entropic security in the classical setting, Def.1 in \cite{dodis2005entropic}]
\label{def:ClassEntrSec}
A probabilistic map is $Y$ called $(t,\qe)$-entropically secure if
\bea{rCl}
	\Hmin(X)\geq t \implies && Y\mbox{ hides all functions of }X \nn \\
    &&\mbox{with leakage }\qe.
\label{defClassicalEntropic}
\eea
\end{definition}
An encryption scheme is $(t,\qe)$-entropically secure if $Y(\cdot)={\sf Enc}(K,\cdot)$ satisfies Def.\,\ref{def:ClassEntrSec}.
The definition is similar to semantic security, but it works with unbounded adversaries and is restricted to high-entropy 
plaintext.

It has been shown that entropic security implies statistical indistinguishability and vice versa, with 
small modifications in the $t$ and $\qe$ parameter.
Indistinguishability is defined by the existence of a variable $G$ such that a ciphertext
resulting from random $X$ (from some distribution) and random $K$
is $\qe$-close to $G$ in terms of statistical distance.  

\begin{definition}[Entropic indistinguishability in the classical setting, Def.2 in \cite{dodis2005entropic}]\footnote{
In \cite{dodis2005entropic} the name `indistinguishability' is used.
Ref.\,\cite{DD2010} sharpened the naming to `entropic indistinguishability' to emphasize the condition on the entropy.
} 
\label{def:ClassEntrDist}
A randomized map $Y$ is called $(t, \qe)$-indistinguishable
if 
\be
    \exists_G\quad \Hmin(X)\geq t \implies \qD(Y(X), G)\leq \qe.
\ee
\end{definition}

\begin{lemma}[Paraphrased from \cite{dodis2005entropic}]
\label{lemma:ClassEquiv}
Let $t\geq 2\log\fr1\qe -5$.
Then $(t-2, 8\qe)$-indistinguishability implies $(t, \qe)$-entropic security for all functions.
\end{lemma}

\begin{lemma}[Claim~2 in \cite{impagliazzo1989recycle}]
\label{lemma:statdistance}
Let $D$ be a distribution on a finite set $\cS$. 
If the collision probability of $D$ is at most $(1+2\varepsilon^2)/|\cS|$, 
then $D$ is at a statistical distance at most $\varepsilon$ from the uniform distribution.
\end{lemma}

\begin{definition}[Quantum encryption scheme]
\label{def:encryption}
A quantum encryption scheme with quantum message space $\cH$,
classical key space $\cK$, and quantum
ciphertext space $\cH'$ consists of a triplet $({\sf Gen}, {\sf Enc}, {\sf Dec})$.
{\sf Gen} is a probabilistic algorithm that outputs a key $k\in\cK$.
The {\sf Enc}$:\cK\times\cD(\cH)\to\cD(\cH')$ is a (possibly randomized) algorithm that takes as input a classical key $k\in\cK$ and a quantum state $\qf\in\cD(\cH)$,
and outputs a quantum state $\qo={\sf Enc}(k,\qf)\in\cD(\cH')$ called the cipherstate. 
${\sf Dec}:\cK\times\cD(\cH')\to\cD(\cH)$ is an algorithm that takes as input a key $k\in\cK$ 
and a state $\qo\in\cD(\cH')$, and outputs
a state ${\sf Dec}(k,\qo)\in\cD(\cH)$.
It must hold that $\forall_{k\in\cK,\qf\in\cD(\cH)}\; {\sf Dec}(k,{\sf Enc}(k,\qf))=\qf$.
\end{definition}

Note that
Def.\,\ref{def:encryption} allows the cipherstate space to be larger than the plaintext space, $\dim\cH' > \dim\cH$. We will be working with the special case where the cipherstate consists of a quantum state of the same dimension as the input, accompanied by classical information.

The effect of the encryption, with the key unknown to the adversary, can be described as a completely positive trace-preserving (CPTP) map
$R:\cD(\cH)\to \cD(\cH')$ as follows,
\be
    R(\qf) = \sum_{k\in\cK} \frac1{|\cK|} {\sf Enc}(k,\qf).
\ee
If the plaintext state is entangled with Eve (E), e.g. the state is $\qf^{\rm AE}$,
then the encryption and decryption act only on the A subspace; we write $R(\qr^{\rm AE})$ for $(R\otimes\one_E)(\qr^{\rm AE})$. 
The definition of entropic security is more involved than in the classical case.

\begin{definition}[Strong entropic security in the quantum setting, Def.4 in \cite{DD2010}]
\label{def:QuantEntrSec}
An encryption system $R$ is called strongly $(t,\qe)$-entropically secure if for all states $\qf^{\rm AE}$ satisfying 
$\Hmin(A|E)_\qf \geq t$, all interpretations $\{(p_i, \qs_i^{\rm AE})\}$ of $\qf^{\rm AE}$, all adversaries $\cA$ and all functions $f$, 
it holds that
\be
	\Big| \Pr[\cA(R(\qs_i^{\rm AE}))=f(i)]- \Pr[\cA(R(\qf^{\rm A})\otimes\qs_i^{E})=f(i)] \Big| \leq \qe.
\ee
\end{definition}
Here `interpretation' means $\qf^{\rm AE}=\sum_i p_i \qs^{\rm AE}_i$.
Def.\,\ref{def:QuantEntrSec} implies another definition of entropic security given in \cite{DD2010}  that contains an adversary $\cA'$ who
gets access only to the own subsystem $\qs_i^{\rm E}$.
Similar to the classical case, the equivalence has been shown between entropic security and entropic indistinguishability in the quantum setting.
\begin{definition}[Entropic indistinguishability in the quantum setting, Def.3 in \cite{DD2010}]
\label{def:QuantEntrInd}
An encryption system $R: \cD(\cH_A)\to\cD(\cH_{A'})$ is called $(t,\qe)$-indistinguishable if 
\bea{rCl}
	\exists_{\qO^{\rm A'}\in\cD(\cH_{A'})} 
	\Hmin(A|E)_\qf &&\geq t \nn\\
	\implies&& 
	\| R({\qf}^{AE})- \qO^{\rm A'}\otimes\qf^E \|_1 \leq \qe.
	\quad
\eea
\end{definition}

\begin{lemma}[Theorem~1 in \cite{DD2010}]
\label{lemma:QuantEquiv}
$(t-1, \qe/2)$-entropic indistinguishability implies strong $(t, \qe)$-entropic security for all functions.
\end{lemma}

\begin{lemma}[Lemma~5.1.3 in \cite{renner2008security}]
\label{lemma:rennerineq}
Let $S$ be a Hermitian operator and $\qs$ a nonnegative operator. 
It holds that
\be
	\big\| S\big\|_1 \leq \sqrt{\Tr(\qs)\Tr(S\qs^{-1/2}S\qs^{-1/2})}.
\ee
\end{lemma}

\subsection{The Quantum One Time Pad (QOTP)}
\label{sec:QOTP}

Let $\cH_2$ denote the Hilbert space of a qubit.
Let $Z$ and $X$ be single-qubit Pauli operators, in the standard basis given by
$Z={1\; \phantom{-}0\choose 0\;-1}$ and $X={0\; 1\choose 1\;0}$. 
For QOTP encryption of one qubit, the key consists of two bits $s,q\in\bits$.
The encryption of a state $\qf\in\cD(\cH_2)$ is given by
$X^s Z^q \qf Z^q X^s$.
Decryption is the same operation as encryption,
but with the order of applying the $Z^q$ and $X^s$ interchanged.

The simplest way to encrypt an $n$-qubit state $\qf\in\cD(\cH_2^{\otimes n})$
is to encrypt each qubit independently.
The QOTP key $\qb$ has length $2n$ and can be expressed as $\qb=s\| q$, with $s,q\in\bits^n$.
In the rest of the paper we will use the following shorthand notation for the  QOTP cipherstate,
\be
    F_\qb(\qf)= U_\qb \qf U_\qb\dagg \quad\mbox{where }
    U_\qb=\bigotimes_{i=1}^n X^{s_i}Z^{q_i}.
\label{defFb}
\ee
Let $\cH_{A'}=\cH_A=\cH_2^{\otimes n}$ and let $\qf^{\rm AE}\in\cD(\cH_A \otimes\cH_E)$ be a bipartite state.
Encryption of the A subsystem is written as
$F_\qb(\qf^{AE})= (U_\qb\otimes\one^E) \qf^{\rm AE}  (U_\qb\dagg\otimes\one^E)$.
For any $\qf^{\rm AE}$ it holds that
\be
	\frac1{2^{2n}}\sum_{\qb\in\bits^{2n}} F_\qb(\qf^{AE}) = \qt^{\rm A'}\otimes\qf^{\rm E},
\label{fullencryption}
\ee
i.e.\;Def.\,\ref{def:QuantEntrInd} is achieved with $t=-n$ and $\qe=0$:
no matter how entangled E is with A, 
from the adversary's point of view
the $A'$ subsystem is fully mixed and entirely decoupled from~E.

\section{Our results on entropically secure classical encryption}
\label{sec:classical}

We present a modification of Ambainis and Smith's second construction in \cite{ambainis2004small}.
The difference lies in the hash function, which in our case consists of the concatenation
of the short key $k$ with an affine function of $k$.
In Section~\ref{sec:proofclassical} we show that
the key length as a function of the security parameter $\qe$ is essentially the same as in \cite{ambainis2004small}.
We comment on the speed gain in Section~\ref{sec:expansion}.

\subsection{The Construction}
\label{sec:constructclassical}

Our construction has message space $\cX=\bits^n$, key space $\cK=\bits^\ell$ (with $\ell\leq n$) and 
ciphertext space $\cC=\bits^{2n}$.
Let $\ql=\max\{\ell,n-\ell\}$.
Let $u\in\bits^\ql$, $v\in\bits^{n-\ql}$.
We define a hash function $h$ as follows,
\be
	h_{uv}(k)\isdef k \| (uk+v)_{\rm lsb}
\label{keyexpansion}
\ee
where $\|$ denotes string concatenation.
The addition and multiplication in the expression $uk+v$ take place in GF$(2^\ql)$;
the subscript `lsb' denotes taking the $n-\ell$ least significant bits in case $\ql>n-\ell$.
Note that $(uk+v)_{\rm lsb}=(uk)_{\rm lsb}+v$. 
The encryption is randomized, with uniformly random strings $u,v$ which become part of the ciphertext,
\be
	{\sf Enc}_{uv}(k,x) = \big( u,v,x \oplus h_{uv}(k) \big).
\label{encclassic}
\ee
Here $\oplus$ stands for bitwise xor.
The decryption is
\be
	{\sf Dec}\big(k, (c_1,c_2,c_3) \big) = c_3\oplus h_{c_1,c_2}(k).
\ee

\subsection{Security Proof}
\label{sec:proofclassical}

\begin{theorem}
\label{theorem:collision}
Our classical encryption scheme described in Section~\ref{sec:constructclassical} satisfies
\begin{eqnarray}
	\pr[{\sf Enc}_{U V}(K,X) &=& {\sf Enc}_{U'V'}(K',X')]
	\nn\\ & \leq &   
    2^{-2n}(1+2^{n-\ell-\sH_2(X)}).
\label{collision}
\end{eqnarray}
\end{theorem}
\underline{\it Proof:}
To start, due to the prepended $u,v$ in (\ref{encclassic})
we get $U'V'=UV$ and we acquire an overall factor $2^{-n}$ from $\pr[UV=U'V']=2^{-n}$.
Next, we need to bound the expression $\pr[X\oplus h_{UV}(K)=X'\oplus h_{UV}(K')]$,
which we can rewrite as
$\pr[X\oplus X' = h_{UV}(K)\oplus h_{UV}(K')]$
$=\sum_{a\in\bits^n}\pr[X\oplus X'=a]\pr[h_{UV}(K)\oplus h_{UV}(K')=a]$.
The $a=0$ term yields a contribution $\pr[K'=K]\pr[X'=X]=2^{-\ell}\pr[X'=X]$.
\bea{rcl}
    &&\pr[U'V'=UV]\pr[X\oplus h_{UV}(K)=X'\oplus h_{UV}(K')] \nn\\
    &&=2^{-n}\Big(2^{-\ell}\pr[X'=X] \nn\\
    &&\quad +\sum_{\stackrel{a\in\bits^n}{a\neq 0}}\pr[X\oplus X'=a]\pr[h_{UV}(K)\oplus h_{UV}(K')=a]\Big) \nn \\
\eea

For the $a\neq 0$ part of the summation we split up $a$ as $a=a_L||a_R$
with $a_L\in\bits^\ell$, $a_R\in\bits^{n-\ell}$ and write
\bea{rcl}
    &&\sum_{\stackrel{a\in\bits^n}{a\neq 0}}\pr[X\oplus X'=a]\pr[h_{UV}(K)\oplus h_{UV}(K')=a] \nn\\ 
    &&= \!\!\!\!\! \sum_{(a_L,a_R)\neq 0}  \!\!\!\!\!   \pr[X\oplus X'=a] \pr[K\oplus K'=a_L]
	\pr[(Ua_L)_{\rm lsb}=a_R] \nn\\ 
    &&=\sum_{\stackrel{a_L\neq 0}{a_R}} \pr[X\oplus X'=a] \pr[K\oplus K'=a_L]  \pr[(Ua_L)_{\rm lsb}=a_R]. \nn \\
\eea
The last equality follows from that fact that $a_L=0$ implies $Ua_L=0$ and hence $a_R=0$, while $(a_L,a_R)=(0,0)$
is not part of the summation; hence $a_L=0$ drops from the summation.
Next, we note that $K\oplus K'$ is uniform on $\bits^\ell$, yielding
$\pr[K\oplus K'=a_L]=2^{-\ell}$ and that $(Ua_L)_{\rm lsb}$ for $a_L\neq 0$
is uniform on $\bits^{n-\ell}$,
yielding
$\pr[(Ua_L)_{\rm lsb}=a_R]=2^{-(n-\ell)}$.
Combining all these results we get
\bea{rCl}
	&& \pr[{\sf Enc}_{UV}(K,X) = {\sf Enc}_{U'V'}(K',X')]
	\nn\\ &&
	=2^{-n}\Big( 2^{-\ell}\pr[X'=X]+2^{-n} \sum_{a_L\neq 0}\sum_{a_R} \pr[X\oplus X'=a] \Big)
	\nn\\ &&
	=2^{-n}\Big( 2^{-\ell}\pr[X'\!=\!X] 
       +2^{-n} [1 \! - \!\sum_{a_R} \pr[X\oplus X' \! = \! 0||a_R]] \Big)
	\nn \\ &&
	\leq 2^{-2n}(1+2^{n-\ell-\sH_2(X)})
\eea
where we have used that $\sH_2(X)=-\log \pr[X'=X]$.
\hfill $\square$ \\

\begin{theorem}
\label{th:mainclassic}
Let $t\geq 2\log\frac1\qe-5$.
Let the key length be set as $\ell = n-t+2\log\frac1\qe-5$.
Then the encryption scheme of Section~\ref{sec:constructclassical}
is $(t,\qe)$-entropically secure.
\end{theorem}

\underline{\it Proof:}
Setting $\ell = n-t+2\log\frac1\qe-5$ in (\ref{collision})
yields a collision probability upper bound of $2^{-2n}(1+2\tilde\qe^2)$,
with $\tilde\qe=8\qe\sqrt{2^{[t-2]-\sH_2(X)}}$.
Thus, according to Lemma~\ref{lemma:statdistance} and Def.\,\ref{def:ClassEntrDist}, our scheme
has $(t-2,8\qe)$ entropic indistinguishability.\footnote{
Note that $\sH_2(X)\geq \Hmin(X)$.
}
Finally we invoke Lemma~\ref{lemma:ClassEquiv} for the implication of $(t,\qe)$-entropic security.
\hfill$\square$

\section{Our results on entropically secure quantum encryption}
\label{sec:quantum}

Our construction for encrypting $n$ qubits is very similar to the classical construction.
The difference lies in the use of the QOTP instead of classical OTP, and in the length of the
expanded key which is now $2n$ instead of~$n$.

\subsection{The Construction}
\label{sec:construction}

The message space is $\cD(\cH_2^{\otimes n})$.
The key space is $\cK=\bits^\ell$.
Let $\ql=\max\{\ell, 2n-\ell\}$.
The ciphertext space is $\bits^{2n-\ell+\ql}\times\cD(\cH_2^{\otimes n})$.
Let $u\in\bits^\ql$ and $v\in\bits^{2n-\ell}$.
We define a hash function $b$ as
\be
    b(k,u,v) = k \| (uk+v)_{\rm lsb}. 
\label{makeb}
\ee
Here the multiplication and addition in $uk+v$ are GF$(2^\ql)$ operations.
The subscript `lsb' (Least Significant Bits) stands for taking the last $2n-\ell$ bits of the string;
in the finite field representation, this corresponds to taking a polynomial in $x$
modulo $x^{2n-\ell}$. 
Instead of $(uk+v)_{\rm lsb}$ we can also write $(uk)_{\rm lsb}+v$.

Let $\qf\in\cD(\cH_2^{\otimes n})$.
The encryption step draws random $u,v$ and outputs $n$ qubits as well as the $u,v$,
\be
    {\sf Enc}_{uv}(k,\qf) = \Big(u,v, F_{b(k,u,v)}(\qf)\Big)
\label{quantumEnc}
\ee
with $F$ the QOTP encryption as defined in (\ref{defFb}) 
and $b(\cdot,\cdot,\cdot)$ as defined by~(\ref{makeb}).
Decryption is essentially the same as encryption,
\be
	{\sf Dec}\big(k, (u,v,\tilde\qf)\big) = F_{b(k,u,v)}(\tilde\qf).
\ee

\subsection{Security Proof}
\label{sec:proof}

Let $\cH_A=\cH_2^{\otimes n}$ and let Eve be entangled with the plaintext state.
The joint state is $\qf^{\rm AE}\in\cD(\cH_A\otimes\cH_E)$.
As discussed in Section~\ref{sec:QOTP}, 
the encryption {\sf Enc} acts only on the `A' subsystem.
As the parameters $u,v$ are public we focus on the quantum part of the ciphertext.
From Eve's point of view, the state after encryption is
\be
    R_{uv}(\qf^{\rm AE}) \isdef \frac1{2^\ell} \sum_{k\in\bits^\ell}  F_{b(k,u,v)}(\qf^{\rm AE}).
\label{defRuv}
\ee

\begin{lemma}
\label{lemma:EvRistau}
It holds that
\be
    \EE_{v} R_{uv}(\qf^{\rm AE})=\qt^{\rm A}\otimes\qf^{\rm E}.
\ee
\end{lemma}

\underline{\it Proof:}
We write
$\EE_{v} R_{uv}(\qf^{\rm AE})=\EE_{kv}F_{b(k,u,v)}(\qf^{\rm AE})$.
Next,
$\EE_{kv}F_{b(k,u,v)}(\qf^{\rm AE})=\EE_{\qb\in\bits^{2n}} F_\qb(\qf^{\rm AE})=\qt^{\rm A}\otimes\qf^{\rm E}$.
The first equality follows from the fact that for any fixed $u$, 
the $k$ and $v$ together can create any string in $\bits^{2n}$ in precisely one way.
The second equality is due to the fact that the QOTP is completely randomizing (\ref{fullencryption}).
\hfill$\square$\\

\begin{lemma}
\label{lemma:sumskkuv}
Let $f$ be any (possibly operator-valued) function acting on $\bits^{2n}$. It holds that
\begin{eqnarray}
    && \EE_{kk'uv} f(b(k,u,v))f(b(k',u,v)) =
    2^{-\ell}\EE_\qb f(\qb)f(\qb)
    \nn\\ && \quad\quad
    + \EE_{\qb \qb'}f(\qb)f(\qb')
    -2^{-\ell}\EE_{kgg'}f(k\| g) f(k\|g').
    \quad\quad\quad
\label{EEkkuv}
\end{eqnarray}
Here $\qb,\qb'\in\bits^{2n}$, and $\EE_\qb$ stands for $2^{-2n} \sum_{\qb}$.
Similarly, $g,g'\in\bits^{2n-\ell}$ and $\EE_g$ stands for $2^{\ell-2n} \sum_g$.
\end{lemma}

\underline{\it Proof:}
We use shorthand notation $g=(uk+v)_{\rm lsb}$, $g'=(uk'+v)_{\rm lsb}$ and we write
$f_x$ instead of $f(x)$. We omit the $\|$ in the expression $k\| g$.
In this notation 
the left hand side of (\ref{EEkkuv}) is $\EE_{kk'uv}f_{kg}f_{k'g'}$.
\bea{rcl}
	&&\EE_{kk'uv}f_{kg}f_{k'g'} \nn \\
    & &=\frac1{|\cK|^2}\sum_k \EE_{uv}f_{kg}f_{kg} + \frac1{|\cK|^2} \sum_{kk': k\neq k'}\EE_{uv}f_{kg}f_{k'g'} 
    \nn\\ 
    & &\stackrel{(a)}{=}
	\frac1{|\cK|}\EE_\qb f_\qb f_\qb + \frac1{|\cK|^2} \sum_{kk': k\neq k'}\EE_{gg'}f_{kg}f_{k'g'}\\ 
	& &=\frac1{|\cK|}\EE_\qb f_\qb f_\qb + \frac1{|\cK|^2} \sum_{kk'}\EE_{gg'} f_{kg}f_{k'g'} 
    -\frac1{|\cK|^2}\sum_k \EE_{gg'}f_{kg}f_{kg'} 
    \nn\\
	& &=\frac1{|\cK|}\EE_\qb f_\qb f_\qb +  \EE_{\qb\qb'} f_\qb f_{\qb'}-\frac1{|\cK|}\EE_k \EE_{gg'}f_{kg}f_{kg'}.\qquad\qquad
\eea
In step (a) we used in the first term that summation over $k\in\bits^\ell$ and $v\in\bits^{2n-\ell}$ exactly corresponds to summation over  
$\qb\in\bits^{2n}$;
in the second term we used that (for $k'\neq k$) averaging  over $(u,v)$
exactly corresponds to averaging over $(g,g')$. 
The latter is obvious in the case $u\in\bits^{2n-\ell}$ ($\ql=2n-\ell$), as the `lsb' in $(uk)_{\rm lsb}$ can be omitted.
In the case $u\in\bits^\ell$ ($\ql=\ell$),
the $(u,v)$-summation covers the $(g.g')$-space exactly an integer number ($2^{2\ell-2n}$) of times.
This is seen as follows.
When the two equations $g=(uk+v)_{\rm lsb}$, $g'=(uk'+v)_{\rm lsb}$
are added, the $v$ disappears and we get
$[u(k+k')]_{\rm lsb}=g+g'$,
which has $2^\ell /2^{2n-\ell}$ solutions~$u$.
Then, at fixed $k,k',g,g',u$ the solution for $v$ is unique.
\hfill$\square$\\

\begin{theorem}
\label{th:normboundq}
Our encryption scheme described in Section~\ref{sec:construction} satisfies
\be
	\EE_{uv} \Big\| R_{uv}(\qf^{\rm AE})-\qt^A\otimes\qf^{\rm E}\Big\|_1 \leq \sqrt{2^{n-\ell-\sH_2(A|E)_\qf}}.
\label{norm1boundq}
\ee
\end{theorem}

\begin{figure*}[b]
\normalsize

\bea{rcl}
\label{wideeqtop}
	&& \EE_{uv}\| R_{uv}(\qf^{\rm AE})-\qt^A\otimes\qf^{\rm E}\|_1 
	\\ && 
    \quad\quad
	\stackrel{\rm Lemma~\ref{lemma:rennerineq}}{\leq} 
	\EE_{uv}\sqrt{\Tr\qs}\sqrt{\Tr[ R_{uv}(\qf^{\rm AE}) \qs^{-\fr12}-\qt^A\otimes\qf^{\rm E} \qs^{-\fr12}]^2} \label{UseRenner}
	\\ && 
    \quad\quad \stackrel{\rm Jensen}{\leq} 
	\sqrt{\Tr\qs}\sqrt{\Tr\EE_{uv}[ R_{uv}(\qf^{\rm AE}) \qs^{-\fr12}-\qt^A\otimes\qf^{\rm E} \qs^{-\fr12}]^2} \label{UseJensen}
	\\ && 
    \quad\quad \stackrel{\rm Lemma~\ref{lemma:EvRistau}}{=} 
	\sqrt{\Tr\qs}\sqrt{\Tr\EE_{uv}[ R_{uv}(\qf^{\rm AE}) \qs^{-\fr12}]^2 -\Tr [\qt^A\otimes\qf^{\rm E} \qs^{-\fr12}]^2}
	\quad \label{UseLemma5}
	\\ && 
    \quad\quad =
	\sqrt{\Tr\qs}\sqrt{\Tr\EE_{kk'uv} F_{b(kuv)}(\qf^{\rm AE}) \qs^{-\fr12} F_{b(k'uv)}(\qf^{\rm AE}) \qs^{-\fr12} 
	-2^{-n}\Tr [\qf^{\rm E} \qs_E^{-\fr12}]^2}.
\label{proofpoint1}
\eea
\hrulefill
\vspace*{4pt}
\end{figure*}
\underline{\it Proof:}
Let $\qs_E$ be a non-negative operator on $\cH_E$.
Let $\qs=\one_A\otimes\qs_E$.
[See (\ref{wideeqtop})--(\ref{proofpoint1}) at the bottom of the page.]
Next, we apply Lemma~\ref{lemma:sumskkuv} to the first expression under the square root in (\ref{proofpoint1}),
taking $f(b(k,u,v))$ $=F_{b(kuv)}(\qf^{\rm AE}) \qs^{-\fr12}$.
This yields
\bea{rcl}
	&&\Tr\EE_{kk'uv} F_{b(kuv)}(\qf^{\rm AE}) \qs^{-\fr12} F_{b(k'uv)}(\qf^{\rm AE}) \qs^{-\fr12} \nn
	\\
    &&=\Tr\Big[2^{-\ell} \EE_\qb F_\qb(\qf^{\rm AE})\qs^{-\fr12} F_\qb(\qf^{\rm AE})\qs^{-\fr12}
    +(\qt^{\rm A}\otimes\qf^{\rm E}\qs_E^{-\fr12})^2 \nn \\
	&&\qquad\qquad -2^{-\ell}\EE_{kgg'} F_{kg}(\qf^{\rm AE}) \qs^{-\fr12} F_{kg'}(\qf^{\rm AE}) \qs^{-\fr12}\Big]
	\\
    &&\leq 2^{-\ell}\Tr \EE_\qb F_\qb(\qf^{\rm AE})\qs^{-\fr12} F_\qb(\qf^{\rm AE})\qs^{-\fr12}+2^{-n}\Tr(\qf^{\rm E}\qs^{-\fr12})^2. \nn \\
\eea
Substitution into (\ref{proofpoint1}), and writing $\Tr\qs=2^n\Tr\qs_E$, gives
\begin{IEEEeqnarray}{l}
	\EE_{uv}\| R_{uv}(\qf^{\rm AE})-\qt^{\rm A}\otimes\qf^{\rm E}\|_1 \nn\\
	\leq\sqrt{2^{n-\ell}}\sqrt{\Tr\qs_E}\sqrt{ \EE_\qb \Tr  F_\qb(\qf^{\rm AE})\qs^{-\fr12} F_\qb(\qf^{\rm AE})\qs^{-\fr12} }\nn\\
    =\sqrt{2^{n-\ell}}\sqrt{\Tr\qs_E}\sqrt{ \EE_\qb \Tr  U_\qb\qf^{\rm AE}U_\qb\dagg \qs^{-\fr12} U_\qb\qf^{\rm AE}U_\qb\dagg\qs^{-\fr12} }\nn\\
    =\sqrt{2^{n-\ell}}\sqrt{\Tr\qs_E}\sqrt{ \EE_\qb \Tr  \qf^{\rm AE}(U_\qb\dagg \qs^{-\fr12} U_\qb)\qf^{\rm AE}(U_\qb\dagg\qs^{-\fr12}U_\qb) }\nn\\
    =\sqrt{2^{n-\ell}}\sqrt{\Tr\qs_E}\sqrt{  \Tr  \qf^{\rm AE} \qs^{-\fr12} \qf^{\rm AE}\qs^{-\fr12} }.
\label{proofpoint2}
\end{IEEEeqnarray}
In (\ref{proofpoint2}) we have used the fact that the $U_\qb$ acts only on the `A' subsystem,
leaving $\qs^{-\frac12}$ unchanged.
Finally, we restrict $\qs_E$ to $\cD(\cH_E)$, so that $\Tr\qs_E=1$, and apply the definition of 
conditional quantum R\'{e}nyi entropy (\ref{condRenyi}) to get the bound
$\Tr  \qf^{\rm AE} \qs^{-\fr12} \qf^{\rm AE}\qs^{-\fr12} \leq 2^{-\sH_2(A|E)_\qf}$.
\hfill$\square$\\

\begin{theorem}
Let the key length be set as $\ell = n-t +2\log\frac1\qe+3$. 
Then the quantum encryption scheme of Section~\ref{sec:construction} is $(t,\qe)$-strongly entropically secure
for all functions. 
\end{theorem}

\underline{\it Proof:}
The ciphertext space is $\bits^{2n-\ell+\ql}\times\cD(\cH_2^{\otimes n})$.
The quantum-classical ciphertext state, entangled with Eve, is
$\qf^{\rm A'E}=\EE_{uv}\ket{uv}\bra{uv}\otimes R_{uv}(\qf^{\rm AE})$.
According to the entropic indistinguishability definition (Def.\,\ref{def:QuantEntrInd}),
the quantity to be bounded is $\| \qf^{\rm A'E}-\qO^{\rm A'}\otimes\qf^{\rm E} \|_1$,
which, after setting $\qO^{\rm A'}=\qt^{\rm A'}$, becomes
$\|\EE_{uv}\ket{uv}\bra{uv}\otimes R_{uv}(\qf^{\rm AE}) -\EE_{uv}\ket{uv}\bra{uv}\otimes \qt^{\rm A}\otimes\qf^{\rm E}\|_1$
=$\EE_{uv}\| R_{uv}(\qf^{\rm AE}) - \qt^{\rm A}\otimes\qf^{\rm E} \|_1$.
This we upper bound with Theorem~\ref{th:normboundq}.
Next, setting $\ell = n-t +2\log\frac1\qe+3$ in (\ref{norm1boundq}) 
yields $\| \qf^{\rm A'E}-\qt^{\rm A'}\otimes\qf^{\rm E} \|_1 \leq$
$\frac\qe2\sqrt{2^{[t-1]-\sH_2(A|E)_\qf}}$.
Hence we have $(t-1,\frac\qe2)$ entropic indistinguishability according to Def.\,\ref{def:QuantEntrInd}.
This implies $(t,\qe)$ strong entropic security according to Lemma~\ref{lemma:QuantEquiv}. 
\hfill$\square$

\vskip2mm

Note that we achieve $(t,\qe)$-entropic {\em indistinguishability}
with key length $n-t+2\log\frac1\qe$. 
For approximate randomization, where the plaintext is unentangled ($t=0$), we thus get the key length $n+2\log\frac1\qe$ as listed in Table~\ref{table:1}.

A special case of quantum encryption is when the plaintext is classical.
Then the quantum encryption scheme typically reduces to a classical scheme that is secure against quantum adversaries.
The QOTP (\ref{defFb}), when applied to classical plaintext bits encoded in the $z$-basis\footnote{
A similar result holds for the $x$ and $y$ basis. 
However, if the $(1,1,1)$ basis is chosen then the result is a quantum encryption scheme for classical
plaintext, with recyclable keys \cite{SdV2017}.
},
has the effect of xor-ing the plaintext with the string $s\in\bits^n$, while the string $q$ does nothing
and can be omitted from the scheme. 

\section{Complexity of the key expansion}
\label{sec:expansion}

We comment on the complexity of our key expansion compared to previous works. 
Complexity is typically quantified as the number of bit-AND operations;
bit-XOR operations are considered to be much cheaper.
Multiplication in GF$(2^\nu)$ has time complexity $\cO(\nu\log \nu)$ 
\cite{AHU1974, cantor1989arithmetical}
whereas GF$(2^\nu)$ addition (subtraction) consists of $\nu$ bit-XOR operations. 
Mateer \cite{mateer2008fast} introduced an improved version of Sch\"{o}nhage's multiplication algorithm \cite{schonhage1977schnelle}. 
If $m$ is of the form $3^\kappa$ and $\kappa$ is  a power of two, then multiplication of two elements in GF($2^{2m}$) requires $ \fr{17}{3}m\log_3 m$ 
bit-AND operations and at least $\fr{52}{5}m\log m \log(\log m) + \fr{3}{2}m\log m - \fr{3}{2}m+\fr{11}{2}\sqrt{m}$ bit-XORs. 
If $\kappa$ is not a power of two, then the number of ANDs slightly increases to $6m\log_3 m$ while the bound on the XORs stays the same.  

\noindent
\underline{Entropically secure encryption}.\\
Dodis and Smith's key expansion for
classical entropically-secure encryption \cite{dodis2005entropic}
makes use of a xor-universal hash function that is implemented by GF($2^n$)-multiplying
the key $k\in\bits^\ell$ times a random string $i\in\bits^n$. 
In contrast, our classical scheme has multiplication in GF$(2^{\lambda})$, and the concatenation 
with $k$ is for free.
For short keys the speedup is modest; only when the key length $\ell$ is a sizeable part of $n$ does the speedup become interesting.

The situation is similar in the general quantum case with entanglement.
The scheme of Desrosiers and Dupuis \cite{DD2010} has a key expansion that
is the same as in \cite{dodis2005entropic} but for string length $2n$ instead of~$n$.
Again, our speedup from GF($2^{2n}$)-multiplication to GF($2^{2n-\ell}$) is modest when the key is short.

\noindent
\underline{Approximate randomization}.\\
The case of approximate randomization corresponds to entropically secure encryption with $t=0$,
i.e. without entanglement between Eve and the plaintext state, but with no further guarantees
about Eve's (lack of) knowledge.
The key size $\ell$ is slightly larger than one-half of the extended length~$2n$.
Our key expansion consists of one multiplication in GF($2^\ell$)
and one addition in GF($2^{2n-\ell}$) or,
since $\ell$ asymptotically almost equals $n$,
roughly speaking one multiplication and one addition in GF($2^n$). 
With Mateer's multiplication for general $\kappa$, this yields a total cost of 
$3 n\log_3 n - 3 n\log_3 2$ ANDs and
$\geq n\log n \{\frac{26}5\log\log\frac n2 +\frac34\}-n\{ \frac{26}5\log\log\frac n2+\frac12 \}$
$+{\cal O}(\sqrt n)$ XORs for our key expansion.

In \cite{ambainis2004small} the $\ell$-bit key $k$ is multiplied by a string 
$\alpha\in\{0,1\}^{2n}$, and the multiplication is in GF($2^{2n}$).
If we write $\alpha=L||R$ and take $\ell\approx n$
then this can be reorganized into the following steps: 
(i) a polynomial multiplication $k\cdot R$ without modular reduction;
(ii) a polynomial multiplication $k\cdot L$ shifted by $n$ positions,
resulting in a polynomial of degree at most $3n$, followed by GF($2^{2n}$)
modular reduction; (iii) addition of the two above contributions.
As we count two XORs per monomial that needs to be reduced\footnote{
With $m=3^\kappa$ it is possible to use the trinomial
$x^{2m}+x^m+1$ as the irreducible polynomial, which allows for
efficient reduction.
If we depart from the $3^\kappa$ form, irreducible polynomials 
of degree 5 may become necessary, which leads to more costly modular reduction.
It has been shown 
 \cite{banegas2019new} that irreducible pentanomials can be chosen such that
no more than {\em three} XORs are required per monomial reduction.
}, we see that
the addition in step (iii) precisely compensates the missing reduction in step (i).
Furthermore, the number of monomials that needs reducing in step (ii) is $n$,
which is the same as in GF($2^n$) multiplication.
Hence the cost of computing $k\cdot\alpha$ equals the cost of 
two GF($2^n$) multiplications.
Since in GF($2^n$) multiplication is much more expensive than addition,
we see that our key expansion is a factor 2 cheaper than~\cite{ambainis2004small}.

\section{Discussion}
\label{sec:discussion}
Our scheme achieves the same shortest key length $\ell=n-t+2\log\frac1\varepsilon$ reported in other studies, 
but with a more efficient key expansion.

It is interesting to note that the security proof in \cite{ambainis2004small} uses Fourier analysis and $\delta$-biased families,
and invokes Cayley graphs for intuition, whereas our proof is more straightforward. 
Furthermore, the security proof in \cite{DD2010} uses an expansion in the Pauli basis, which we do not need.

We note that in the classical case our scheme always needs public randomness that is as large as the plaintext.
In the quantum case, however, the size of the public randomness is not the same in the two cases that we distinguish
($\ell<n$ versus $\ell> n$).
It is left for future work to see if this can be improved.

All the definitions of entropic indistinguishability and entropic security
put a condition on the {\em min-entropy} of the plaintext, 
whereas all the security proofs yield expressions that contain the {\em collision entropy}.
It may be advantageous to work with modified definitions that put a condition on the
collision entropy, since that is more easily met.

\section*{Acknowledgments}
We thank Tanja Lange and Dan Bernstein for discussions on multiplication complexity.

\bibliographystyle{IEEEtran}
\bibliography{Entropic}

\end{document}